\documentclass[prb,reprint,letterpaper,numerical]{revtex4-1}
\usepackage{color}
\usepackage{geometry}
\usepackage{graphicx}
\usepackage{amssymb}
\usepackage{bm}
\usepackage{amsmath}

\usepackage{ulem}

\begin{document}

\title{When does TMAO fold a polymer chain and urea unfold it?}

\author{Jagannath Mondal} 
\affiliation{Department of Chemistry, Columbia University, New York, NY 10027}

\author{Guillaume Stirnemann}
\affiliation{Department of Chemistry, Columbia University, New York, NY 10027}

\author{B. J. Berne}
\email{bb8@columbia.edu}
\affiliation{Department of Chemistry, Columbia University, New York, NY 10027}

\date{\today}
\begin{abstract} 

  Longstanding mechanistic questions about the role of protecting
  osmolyte trimethylamine \textit{N}-oxide (TMAO) which favors protein
  folding and the denaturing osmolyte urea are addressed by studying
  their effects on the folding of uncharged polymer chains. Using
  atomistic molecular dynamics simulations, we show that 1-M TMAO and
  7-M urea solutions act dramatically differently on these model
  polymer chains. Their behaviors are sensitive to the strength of the
  attractive dispersion interactions of the chain with its
  environment: when these dispersion interactions are high enough,
  TMAO suppresses the formation of extended conformations of the
  hydrophobic polymer as compared to water, while urea promotes
  formation of extended conformations. Similar trends are observed
  experimentally on real protein systems. Quite surprisingly, we find
  that \textit{both} protecting and denaturing osmolytes strongly
  interact with the polymer, seemingly in contrast with existing
  explanations of the osmolyte effect on proteins. We show that what
  really matters for a protective osmolyte is its effective depletion
  as the polymer conformation changes, which leads to a negative
  change in the preferential binding coefficient. For TMAO, there is a
  much more favorable free energy of insertion of a single osmolyte
  near collapsed conformations of the polymer than near extended
  conformations. By contrast, urea is preferentially stabilized next
  to the extended conformation and thus has a denaturing effect.

\end{abstract}

\maketitle

\section{Introduction}

Osmolytes are small cosolutes found endogenously to protect cells
against osmotic stress\cite{YANCEY:1982fk}. However, they can have
profound effects on protein
stability\cite{WYMAN:1964vn,Tanford:1969kx,Timasheff:2002kx,Parsegian:2000uq,bolen,Courtenay:2000uq,Canchi:2013fk}. While
some of them are denaturants (e.g. urea), others like trimethylamine
\textit{N}-oxide (TMAO) act as protecting osmolytes in vivo: in
denaturing conditions, they bias the protein structure toward the
folded conformation
\cite{Street:2006uq,Wang:1997kx,YANCEY:1982fk,LIN:1994vn}. They
are thus referred to as chemical chaperones. Hence, TMAO is used by
deep-sea organisms to counteract the deleterious effect of pressure
and by sharks or skates to compensate for their relatively high
concentrations of the denaturing urea.  Most
interestingly, the protein folding propensity of TMAO has been used
experimentally to study the mechanisms involved in protein misfolding
diseases, including e.g. prion protein
\cite{Tatzelt:1996ys,Nandi:2006zr}, tau protein
\cite{Tseng:1998ve,Scaramozzino:2006ly} (Alzheimer disease) and
alpha-synuclein \cite{Uversky:2001bh} (involved in numerous
neurodegenerative diseases); chemical chaperones like TMAO even appear
promising as therapeutics \cite{Morello:2000cr}, even though it was
recently found to be related to an increased risk of cardiovascular
diseases in humans \cite{Tang:2013fk}. TMAO is active in endogenous
systems for concentrations as low as 200 mM
\cite{Tatzelt:1996ys,YANCEY:1982fk}. Experiments in vitro on
alpha-synuclein, an intrinsically disordered protein, have given
evidence that this protective effect increases with concentration.
\cite{Uversky:2001bh}

TMAO is a small amphiphile (chemical formula: (CH$_3$)$_3$NO) consisting of a small hydrophilic group
(N$^+$O$^-$) and a bulky hydrophobic part (3 methyl groups). Several
mechanisms have been invoked to explain the folding propensity of
TMAO. In a first scenario, TMAO would enhance water structure and
hydrogen-bond (HB) strength, which would indirectly affect the
equilibrium between the folded and the unfolded conformations of a
protein.\cite{Bennion:2004fv,Sharp:2001nx,Hunger:2012fk} However, this mechanism has
been challenged by several studies, mainly based on molecular dynamics
(MD) simulations, where no significant alteration of water structure
was found in aqueous solutions of TMAO
\cite{Athawale:2005kl,Cho:2011fv,Hu:2010tg}. These results may not be
surprising since TMAO can only accept 2 to 3 strong HBs at its
hydrophilic head, which represents less than $10\%$ of its hydration
water HB population \cite{Stirnemann:2010ij}.

Other studies have suggested that direct interactions, or especially
the lack thereof, between TMAO and the protein backbone could cause
the osmolyte effect. In particular, thermodynamic measurements have
highlighted the importance of the interactions between TMAO and the
protein backbone and side-chains
\cite{Street:2006uq,Wang:1997kx}. TMAO has favorable interactions with
some protein side-chains, especially the positively charged groups
that can interact with the O$^-$ of TMAO. In contrast, interactions
with the protein backbone, in particular with the amide NH, are
entropically unfavorable.\cite{Cho:2011fv} If these unfavorable
interactions were to dominate, TMAO would be depleted from the protein
surface. It has been suggested that the resulting concentration
gradient in the TMAO could lead to an osmotic pressure favoring the
folded conformation with respect to the unfolded one.
\cite{Cho:2011fv} Recently, Garcia and co-workers have combined
\cite{garcia} computer simulation with the experimental osmotic
pressure measurements and have suggested a mechanism where there is
preferential exclusion of TMAO from protein surfaces due to repulsive
self-interaction in the solvation shell.  Others have argued that
osmotic pressure itself cannot explain this phenomenon, and that
``water mediated interactions" between the osmolyte and the protein
could also play a role.\cite{Hu:2010tg}

However, a clear unifying scenario has not yet emerged from the study
of the effect of TMAO on proteins. One of the reasons might be the
presence of amino acids with different chemical properties which might
complicate the role of TMAO as a structure enhancer. Thus instead of
struggling with different amino acids, a successful strategy can be to
use a simple polymeric chain whose hydrophobicity can be tuned.  There
have already been studies of the action of TMAO on purely hydrophobic
chains\cite{Athawale:2005kl}, but even for such a simple system a
consensus has not yet been achieved. A good illustration of the lack
of consensus is the opposite conclusions reached in two different
investigations. Based on simulations of a small hydrophobic solutes
and of hydrophobic chains, one of these studies suggested that TMAO
has a negligible effect on the hydrophobic
interactions.\cite{Athawale:2005kl} In contrast other studies suggest
that TMAO destroys hydrophobic interactions
\cite{Paul:2007kl}.
  
To investigate the molecular mechanism of TMAO's role as a protective
osmolyte, it is interesting to compare it with the effect of urea (chemical formula: (NH$_2$)$_2$CO)
solutions that lead to the opposite behavior --- unfolding of the
protein. For example Pettitt and co-workers\cite{Pettitt:2011ja} have
recently explored the conformational preferences of decaalanine in
TMAO and urea solutions using free energy perturbation
techniques. Their analysis, based on the decomposition of the transfer
free energy, suggests the differences in the behavior of peptide in
the two different solutions arises mainly from differences in the
relative importances of van der Waals and electrostatic interactions:
urea denaturation is dominated by van der Waals attractions whereas TMAO exterts
its effect by causing unfavorable electrostatic interactions. In this
contribution we extend our previous work on urea \cite{Zangi:2009mi}
by using similar systems and methodologies and apply it to contrast
the respective role of a denaturing osmolyte (urea) and a protective
osmolyte (TMAO) on uncharged chains in water. We focus on the
mechanisms by which these two osmolytes produce opposite actions on
the conformations of this Lennard-Jones chain. Following most of the
previous simulation studies, we used concentrations higher than that
found in vivo to enhance the osmolyte influence on protein
stability. Hence we chose a concentration of 7 M for urea (consistent
with Ref.~\cite{Zangi:2009mi}) and a concentration of 1 M for
TMAO. Both concentrations are widely used to study the effects of the
respective osmolytes {\it in vitro} and {\it in silico}. Our study
shows that while acting on the same chain, TMAO stabilizes the
collapsed conformations of the chain while urea destabilizes the
collapsed conformations, and the simulations later allow us to offer a
molecular explanation to these different behaviors.  The paper is
organized as follows: the simulation model and methods are described
in section II, results are presented in section III, and some
conclusions are presented in section IV.

\section{Simulation model and methods}

{\it System and forcefields} Our system consists of a 32-bead polymer
solvated in various aqueous solutions. The polymer is uncharged and
the beads only interacts with their environment via Lennard-Jones (LJ)
potentials. While the bead radius is fixed ($\sigma_b$=0.4 nm) the
hydrophobic character of the chain can be tuned by varying the energy
parameter $\epsilon_b$. Following a previous study,\cite{Zangi:2009mi}
four values were employed ($\epsilon_b= 0.4, 0.6, 0.8, 1.0$~kJ/mol)
even if most of the current work was performed on the
$\epsilon_b=1.0$~kJ/mol polymer. Among the chain, 1-4 interactions
were removed; parameters for the 1-2 (bonds) and 1-3 (angles)
interactions can be found elsewhere.\cite{Zangi:2009mi} For water
molecules, we used the SPC/E model\cite{Berendsen:JPhysChem:1987},
while urea interacts through the OPLS/AA forcefield \cite{opls} and
TMAO through the forcefield developed by Kast et al's\cite {kast}. The
geometric combining rules were used in the cross interactions for
$\epsilon$ and arithmetic combination rules were used for $\sigma$.
Three systems were simulated. The system of pure aquous solution was
composed of the polymer solvated by 4092 water molecules. The system
of 1 M TMAO solution was composed of the 32-bead polymeric chain, 79
TMAO molecules and 4013 water molecules. On the other hand, the system
of 7 M urea solution was composed of the 32-bead polymeric chain, 500
urea molecules and 2727 water molecules. The box size was close to $5\times5\times5$~nm$^3$ in all cases. We have also repeated our
simulation for TMAO at $\epsilon_b=1.0$~kJ/mol using a different
forcefield (called herein the ``osmotic model") recently proposed by
Garcia and coworkers\cite{garcia} and we found it to follow the same
qualitative trends ({\it c.f.} the Appendix).

{\it Equilibrium simulations} All simulations were performed using
Gromacs 4.5.4 software \cite{gromacs4}. In a bid to sample different
polymer conformations in different osmolyte solutions, unrestrained
equilibrium MD simulations of the polymer chain were performed in pure
water, 1 M TMAO and 7 M urea solutions. In order to avoid any bias, an
extended configuration of the polymeric chain was used as an initial
configuration in pure water and in the aqueous solution of 1 M TMAO
while a collapsed configuration of the polymer was used as an initial
configuration for the simulation in aqueous solution of 7 M urea. The
initial extended configuration was an all-trans configuration of the
polymer while the collapsed configuration was picked from the
simulation of the polymer in water.  Each of the systems was first
energy-minimized using a steepest-descent algorithm and then subjected
to 100 ns of production run in NPT ensemble. The Nose-Hoover
thermostat was used for maintaining the average temperature at 300 K
and the Parinello-Rahman barostat was used for maintaining the average
pressure at 1 bar. For all three aqueous solutions, unrestrained
simulations were repeated for the four values of the LJ energy
parameter ($\epsilon_b$) for the polymer beads.

{\it Potentials of mean-force} We determined the free energy landscape
(or potential of mean force [PMF]) of the 32-bead LJ chains along one
or several collective reaction coordinates in different solutions by
performing umbrella sampling simulations. We chose as reaction
coordinate the polymer radius of gyration $R_g$ in pure water, 1 M
TMAO and 7 M urea. We employed the PLUMED extension of
Gromacs\cite{plumed}. The value of $R_g$ ranged from 0.4 nm to 1.2 nm
at a spacing of 0.05 nm between adjacent windows.  Restraining
harmonic force constants of 7000 kJ/mol/nm$^2$ were used in the
umbrella potential in all positions to ensure a Gaussian distribution
of the reaction coordinate around each desired value of the reaction
coordinate.  Finally, we used the Weighted Histogram Analysis Method
(WHAM).  \cite{kumarBSKR92,grossfield} to generate unbiased histograms
and corresponding free energies. As described later we also compute
the joint probability distribiution of $R_g$ and the end-to-end
distance and the corresponding potential of mean force as a function
of these two variables.

{\it Preferential interaction} We employed two parameters to measure
the affinity of the cosolvent (urea or TMAO) for the polymer. First,
the local-bulk partition coefficient $K_p$ was calculated, where
as\cite{Courtenay:2000uq}
\begin{equation}
\label{eq:pref_bind} 
K_p =\frac
{\langle n_{s} \rangle}{\langle n_{w}\rangle} \cdot \frac {N^{tot}_{w}}{N^{tot}_{s}}.
\end{equation}
Here $\langle n_X \rangle$ is the average number of molecules of type X bound to
polymer and $N^{tot}_X$ is the total number of molecules of type X in the
system (where $X=s$ stands for the cosolvent (urea or TMAO) and
$X=w$ stands for water). $K_p$ is intensive and reflects the affinity
of the cosolvent for the polymer regardless of the exposed surface
area of the polymer. The other parameter is the
experimentally-relevant preferential binding
coefficient,\cite{WYMAN:1964vn,Tanford:1969kx,Courtenay:2000uq,Shukla:2009fk,garcia}
\begin{equation} 
  \Gamma = \left \langle n_{s} - \frac {N^{tot}_{s}-n_s}{N^{tot}_{w}-n_w}\cdot n_{w} \right \rangle,
\label{eq:pref_bind2}
\end{equation}
which is extensive (i.e., it depends on the size of the hydration
shell). To determine the dependence on the proximal cut-off distance
for the counting of molecules around the polymer, we computed the
value of both quantities $K_p$ and $\Gamma$ as a function of distance
from the polymer (i.e., by examining the explicit distance dependance
of $n_s(r)$ and $n_w(r)$), which is defined as the shortest distance
between the central atom of the solvent molecule (O for water, N for
TMAO and C for urea) and any polymer bead. Additional simulations were
performed as follows. We froze 5 representative configurations of the
the polymer either in the collapsed or the extended state for each
osmolyte solution, and then propagated each of these simulations for
15~ns (total simulation length of 75 ns for each polymer
configuration). In each case, both $K_p$ and $\Gamma$ were averaged
over this ensemble of trajectories, and standard deviations were
obtained via block averaging.

Finally, we used the Free Energy Perturbation (FEP) technique to
compute the transfer free-energy (chemical potential) for inserting a
{\it single} TMAO (or urea or water molecule) from bulk solution into
the first solvation shell of particular conformations of polymer in 1
M TMAO (or 7 M urea) where the polymer conformation was fixed in
either a collapsed or an extended conformation.  For these
calculations the initial configurations were taken from a
representative snapshot of the prior umbrella sampling windows and the
TMAO (or urea or water) molecule was grown in presence of other TMAO
(or urea or water) molecules in solution. The interactions of the
molecule being inserted were slowly turned on in two stages: in the
first stage only the van der Waals interactions were turned on, and in
the second stage the electrostatic interactions were turned
on. Thermodynamic integration gives these two contributions to the
transfer free energy.  The difference between the free energy for the
insertion proximate to the polymer and the insertion in bulk gives the
required transfer free energies. Finally, since the choice of the
position near the polymer where the osmolyte and solvent molecules is
grown is arbitrary, we repeated such calculations at several positions
in each case: 5 sites for TMAO and urea and 3 sites for water
  near each of collapsed and extended conformations and 2 sites of
  each of them in bulk media. They were finally averaged and the
standard deviations were estimated.

\section{Results and discussion}

\subsection{Osmolyte effect on folding equilibria} 

\begin{figure} [h!] \centering
\includegraphics[clip,width=1.0\columnwidth,keepaspectratio]{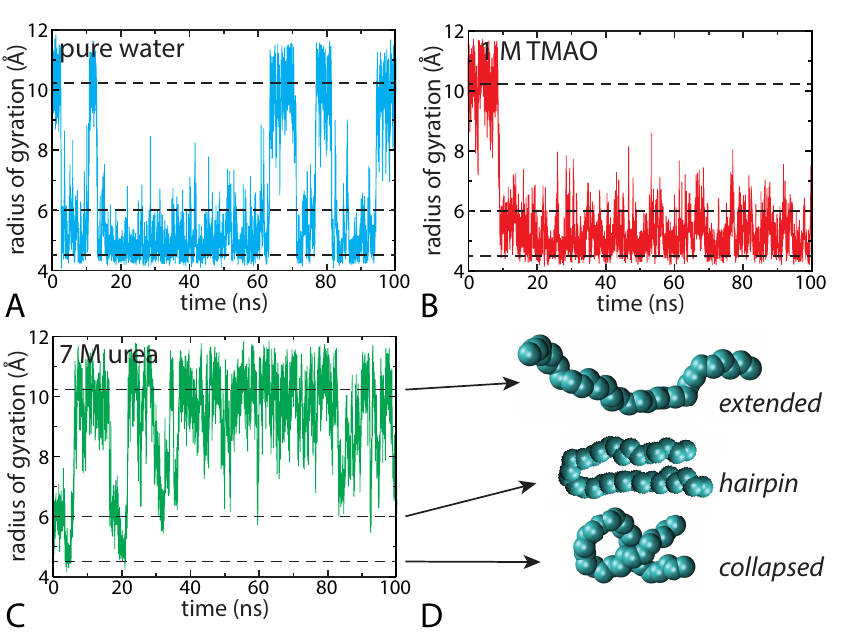}
\caption{Time profile of polymer's radius of gyration for
  $\epsilon_b= 1.0$~kJ/mol as obtained from unrestrained
  simulations in different aqueous solutions: (A) in water, (B) in 1 M
  TMAO solution and (C) in 7 M urea solution. The horizontal dashed lines corresponding to $R_g$ = 4.5, 6.0 and 10.2 \AA~represent the
  most probable values of the radius of gyration for the collapsed
  (folded), hairpin-like and extended (unfolded) conformations
  respectively, depicted in (D).}
\label{rg_profile}
\end{figure}

We first verify that the effects of TMAO and urea on an uncharged
Lennard-Jones polymer chain are those observed in protein systems,
i.e. that they act respectively as protective and denaturing
osmolytes. Towards this end, a reasonable approach is to follow the
time-evolution of an order parameter describing the polymer
conformation, like its radius of gyration which will allow us to
distinguish between the collapsed and extended configurations. Such
profiles are shown for a polymer with a bead interaction parameter
$\epsilon_b= 1.0$ kJ/mol in Figure~\ref{rg_profile}. For these
unrestrained simulations, the initial configurations in each system
(water, 1 M TMAO and 7 M urea) were chosen anticipating the effect of
this aqueous solution on the polymer conformational equilibrium. Thus,
simulations were started from an unfolded, state in water and TMAO
solutions ($R_g=11.5$~\AA), and started from a collapsed configuration
($R_g=4.5$~\AA) in the urea solution. Each of these simulations were
then propagated for 100~ns.

 Figure ~\ref{rg_profile} shows that the polymeric chain behaves
  very differently in the three environments. In water (Figure
  ~\ref{rg_profile}A), the initially-extended polymer quickly
  collapses and then fluctuates between the collapsed and the extended
  configurations. In the TMAO solution (Figure ~\ref{rg_profile}B),
  the polymer collapses and remains compact for the whole 100-ns
  timescale: the extended configuration sampled in pure water is not
  observed in TMAO on the time scale of the simulation. In contrast,
  the polymer in urea unfolds, (Figure~\ref{rg_profile}C) but very
  occasionally revisits more compact states like the hairpin at
  $R_g\approx 0.6$~nm, and very rarely visits the most compact states
  seen in water.

  We thus see that for $\epsilon_b$=1.0~kJ/mol, TMAO and urea act
  respectively as protective and a denaturing osmolytes with respect
  to the hydrophobic chain. This was already observed for
  urea\cite{Zangi:2009mi} albeit for a different water model
  (TIP4P\cite{tip4p}). It is remarkable that these osmolytes have
  similar effects on the hydrophobic chain as they do on real
  proteins.  In the following, we aim to better understand this
  interesting behavior.

\begin{figure}[t] \centering
\includegraphics[clip,width=1.0\columnwidth,keepaspectratio]{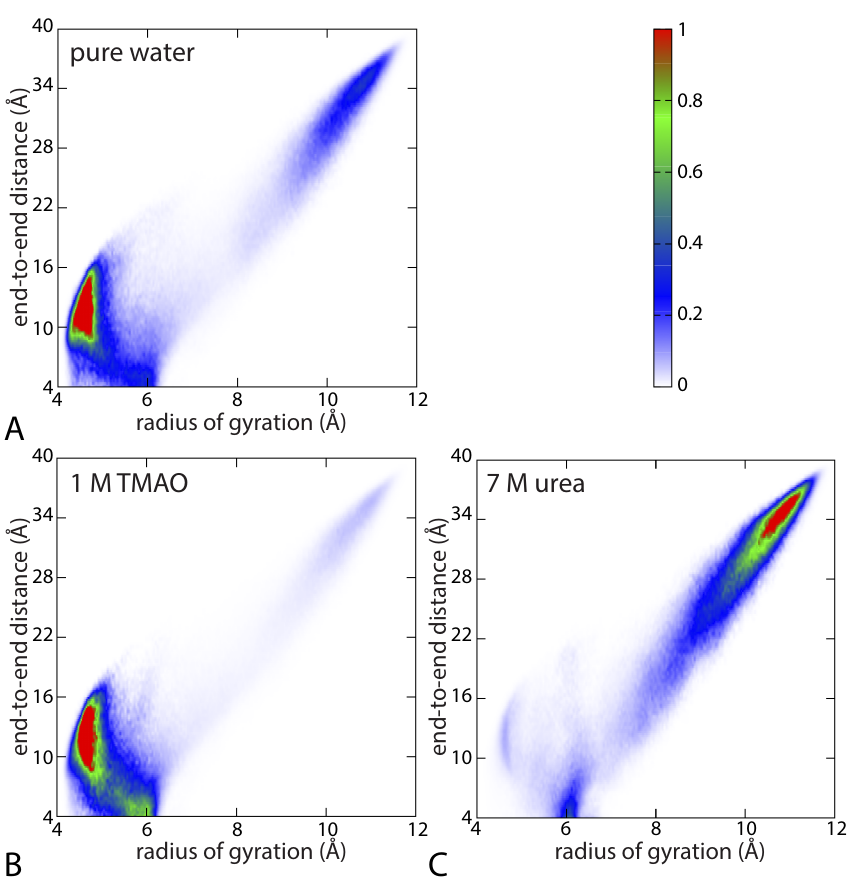}
\caption{Comparison of joint-probability distribution of polymer chain
  (with $\epsilon_b=1$ kJ/mol) along radius of gyration and end-to-end
  distance (A) in water, (B) in 1 M TMAO and (C) in 7 M urea, obtained
  from umbrella sampling simulations.}
\label{2d_prob}
\end{figure}

The semi-compact hairpin configuration of the chain ($R_g\approx
0.6$~nm) observed in the urea solution\cite{Zangi:2009mi} (see Figure
~\ref{rg_profile}C) and to some extent in water and TMAO solution
(Figure ~\ref{rg_profile}A and Figure ~\ref{rg_profile}B) can be
better understood by considering a two dimensional collective
coordinate consisting of the radius of gyration and the end-to-end
distance of the chain. Figure~\ref{2d_prob} shows the joint
probability $P(L,R_g)$ of finding a polymer ($\epsilon_b$=1.0~kJ/mol)
with end-to-end distance $L$ and radius of gyration $R_g$, for each of
the systems shown in in Figure ~\ref{rg_profile}. To avoid possible
biases due to limited sampling in unperturbed simulations, the
probability distribution of $R_g$ is first recovered from the PMF
obtained via umbrella-sampling simulations. In each window, we later
estimate the conditional probability $P(L|R_g)$ of finding  $L$
\textit{given} $R_g$. The joint probability is finally 
recovered using the relation
  \begin{equation}
\label{joint_prob} P(L,R_g) =P(L|R_g)P(R_g).
\end{equation}

The distributions shown in Figure~\ref{2d_prob} confirm our previous
findings: in urea the extended configurations are significantly more
populated than in water whereas in TMAO they are essentially
absent. Moreover, the hairpin state observed at $R_g\approx 0.6$~nm
and small $L$ in the 2D plots, although more prominent in urea
solution, also appear in TMAO solution and in pure water.  In
contrast, the collapsed state around $R_g~\approx~0.45$~nm corresponds
to higher $L$, showing that in the two order parameters are largely
decoupled in this region of the distribution. Unfolded configurations
correspond to high values of both $L$ and $R_g$ and give rise to
distributions elongated along the diagonal in the 2D plots and only
urea has a strong peak there.
 
\begin{figure}[t] \centering
     \includegraphics[clip,width=1.0\columnwidth,keepaspectratio]{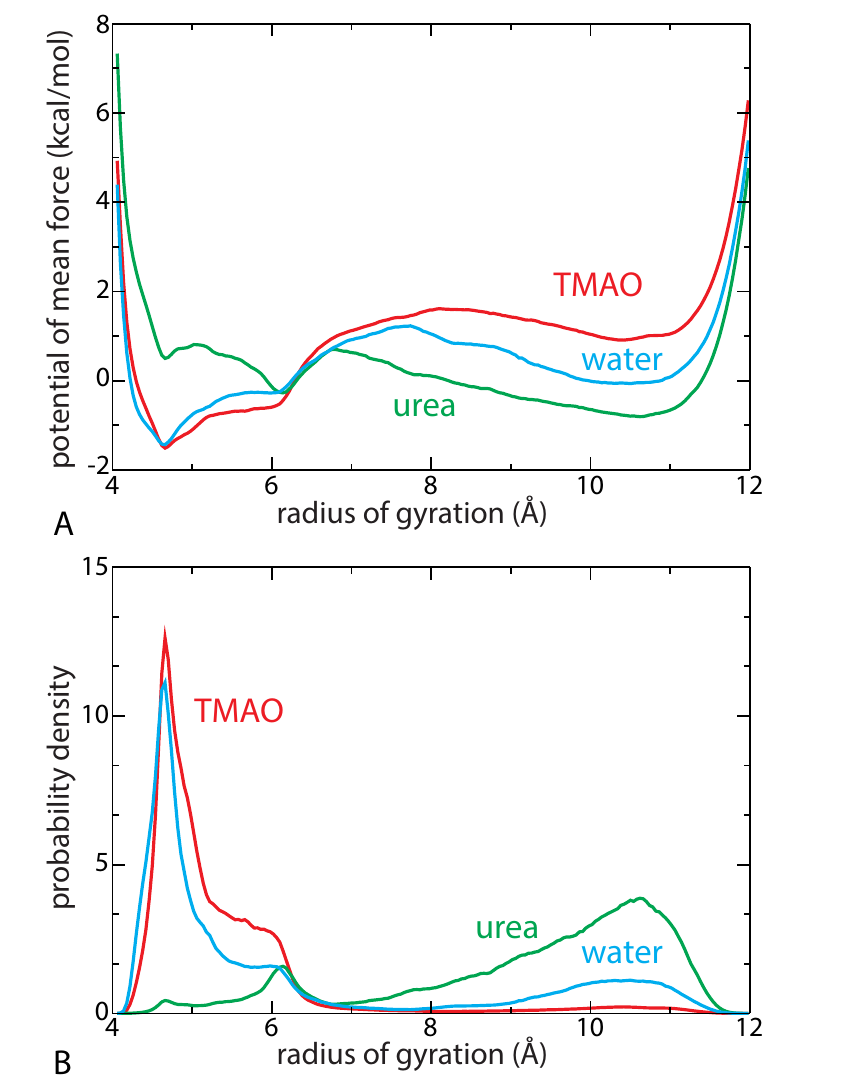}
     \caption{Potential of mean force along the radius of gyration (A)
       and the corresponding probability distributions (B) in the
       three aqueous solutions (water, blue; TMAO, red and urea,
       green) for polymer chains with $\epsilon_b=1.0$ kJ/mol. The
       PMFs $W(R_g)$ are normalized so that $\int_0^\infty
       \exp(-W(r)/k_BT)dr=1$.} \label{pmf_all}
   \label{fig:pmf}
\end{figure}

 The potentials of mean force $W(R_g)$ as a function of the polymer
  radius of gyration $R_g$, obtained via umbrella sampling,
  corresponding probability distributions $\exp[-\beta W(R_g)]$ (for
  $\epsilon_b$=1.0 kJ/mol) and for the three aqueous solutions are
  shown in Figure~\ref{pmf_all}A and B respectively (and in the Figure ~\ref{pmf_all_no_jacobi}  for other values of $\epsilon_b$).  These correspond to the
  projections of the joint-probability distribution onto the radius of
  gyration axis. These provide a more reliable and quantitative
  description of the polymer conformational equilibrium than
  simulations  based on unrestrained MD
  trajectories (Figure~\ref{rg_profile}) because those would require much longer runs (as later
  illustrated by the free-energy barriers of 2 to 4~kcal/mol between
  states).

 As can be seen in Figure~\ref{fig:pmf}, the unfolded state
  in TMAO solution gets destabilized with respect to
  pure water, whereas the collapsed state is not
  dramatically affected. Quite remarkably, the unfolded state  is
  almost totally suppressed for TMAO and its collapsed state is more
  compact for this case of $\epsilon_b=1.0$ kJ/mol.

 In previous molecular dynamics simulations of a similar system,
  \cite{Athawale:2005kl} it was found that that TMAO has little affect
  on the conformational equilibrium of a hydrophobic polymer chain. In
  that study a lower polymer bead parameter of $\epsilon_b=0.6$~kJ/mol
  was used than in our simulations. It is therefore of importance to
  determine the effect of smaller $\epsilon_b$ on the behavior of
  TMAO, as was done earlier for urea solutions\cite{Zangi:2009mi}. In
  the Appendix and Figure~\ref{pmf_all_no_jacobi} that as the hydrophobicity of the the chain is
  increased or equivalently as $\epsilon_b$ is decreased the chain
  responds differently to TMAO.  effect of 1 M TMAO
  solution is thus very sensitive to the value of $\epsilon_b$. We
  find that the protecting role of TMAO is very weak when the
  chain is strongly hydrophobic (low values of $\epsilon_b=$ 0.4 and
  0.6 kJ/mol), in agreement with the conclusions of the previous
  study.\cite{Athawale:2005kl} However TMAO's protecting role becomes
  much more prominent for larger values $\epsilon_b=$ 0.8 and
  1.0~kJ/mol.

 As previously suggested using a different water model
\cite{Zangi:2009mi}, the response of the polymer to urea on
decreasing $\epsilon_b$ is radically different. The trend is clearly
opposite to that found in water or in TMAO solution: the unfolded
state gets stabilized while the collapsed state population
progressively disappears, and a significant fraction of the population
is found in the hairpin state. Urea therefore exhibits a typical
denaturing effect. In strong contrast with TMAO and in agreement with
a previous study\cite{Zangi:2009mi}, we show in the Figure~\ref{pmf_all_no_jacobi} that
urea readily denatures hydrophobic polymers (e.g.
$\epsilon_b=$~0.4~kJ/mol) yet its  denaturing effect becomes more
prominent as $\epsilon_b$ increases.

To assess the robustness of our results, we also repeated our
simulations $\epsilon_b=1.0$~kJ/mol using a different water
model\cite{tip4p} and found very similar results (see Appendix and
Figure~\ref{PMF_water_model} ). Although the force-field we have employed for TMAO has
been widely employed in the past and clearly behaves as a protective
osmolyte, \cite{Cho:2011fv,Hu:2010tg,Pettitt:2011ja} it has recently
been criticized because it underestimates osmotic pressure at high
solute concentrations.\cite{garcia} We have repeated our simulations
at 1 M using Garcia et al's modified ``osmotic") version of this
force-field.\cite{garcia} As shown in Figure~\ref{PMF_Garcia}, its effect on the
polymer chain nonetheless differs very little from what we observed
using the aforementioned forcefield.

Our simulations show that TMAO acts as a protective osmolyte and urea
as a denaturant on the polymer chain for $\epsilon_b=1.0$~kJ/mol. TMAO
thus acts on this chain similarly to the way it acts on many proteins
as found experimentally~\cite{bolen}, showing that its effect extends
to purely uncharged polymer chains of moderate hydrophobicity.  TMAO's
ability to act as a protecting osmolyte depends on the properties of
the polymer it is acting on as is shown by its sensitivity to the
value of the polymer-bead $\epsilon_b$: TMAO seems to have little
effect on strongly hydrophobic chains. Urea, on the other hand still
denatures them. In the following, we aim to better understand the
effect of both osmolytes on the polymer chain with
$\epsilon_b=1.0$~kJ/mol.

\subsection{Molecular Mechanism of Osmolyte-induced (un)folding} 

\subsubsection{Interpretation based on preferential binding and
  chemical potential of osmolytes}

It has been suggested in the literature that denaturants exhibit
preferential interaction with protein surfaces while protective
osmolytes are preferentially excluded from the
surface\cite{Timasheff:2002kx,Parsegian:2000uq,bolen,Courtenay:2000uq,Canchi:2013fk}
because of unfavorable interactions
\cite{bolen,Canchi:2013fk}. Therefore such behavior should be observed
in both the local-bulk partition coefficient $K_p$
(Eq.~\ref{eq:pref_bind}) and the preferential binding coefficient
$\Gamma$ (Eq.~\ref{eq:pref_bind2}).
  
\begin{figure}[h] \centering
\includegraphics[clip,width=1.0\columnwidth,keepaspectratio]{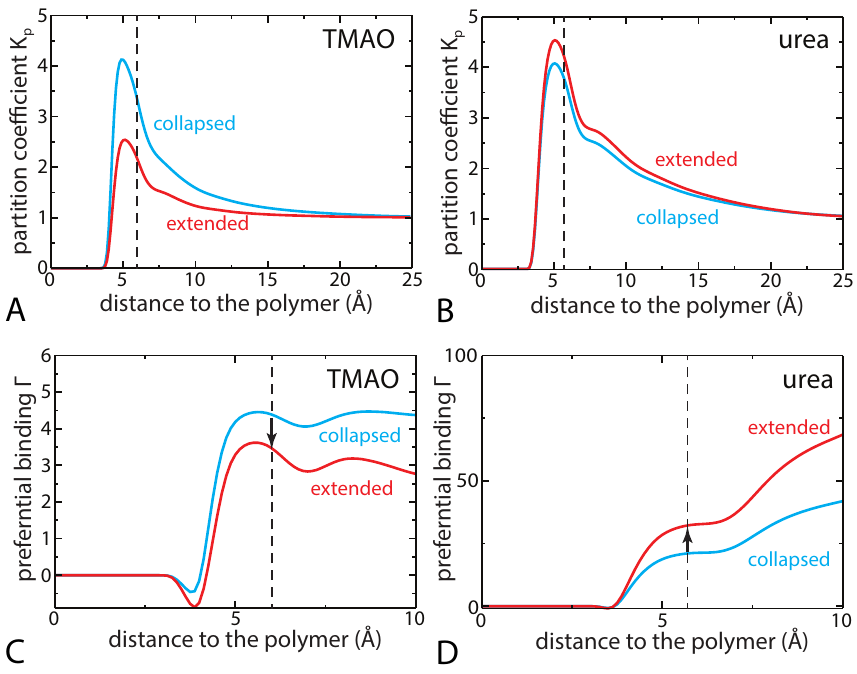}
\caption{Preferential binding constants of TMAO in 1 M TMAO (red) and
  urea in 7 M urea (green) as a function to the distance to the
  polymer (for chains with $\epsilon_b=1.0$~kJ/mol), frozen in an
  collapsed (black) or extended (red) configurations extended
  configuration. Vertical dashed lines indicate the position of the
  polymer first solvation shell in each case (6~\AA~for TMAO and
  5.7~\AA~for urea), and the black arrows in C and D represent the
  relevant value of $\Delta\Gamma$.}
\label{fig:pref_binding}
\end{figure}

In Figure~\ref{fig:pref_binding}A-B we compare the dependence of $K_p$
on polymer conformation as a function of distance from polymer in both
TMAO and urea solutions. If the local domain is defined as the
polymer's first solvation shell, then the representatives values for
$K_p$ (and later $\Gamma$) should be taken near $\approx 6$~\AA
~(dashed line), which corresponds to the the first minima of the
radial distribution functions between the polymer and cosolvent
molecules (urea or TMAO). As expected, urea molecules accumulate next
to both collapsed and extended states of the polymer, leading to
$K_p>1$ (Figure~\ref{fig:pref_binding}B). $K_p$ is little sensitive to
the polymer conformation but it is always slightly higher in the
extended state. The local-bulk partition coefficient of TMAO is also
\textit{greater} than 1 (Figure~\ref{fig:pref_binding}A), implying
that TMAO also \textit{binds} to the polymer surface. This result
contradicts the popular view that protective osmolytes are believed to
be preferentially excluded from the protein
surface\cite{Timasheff:2002kx,Parsegian:2000uq,bolen,Courtenay:2000uq,Canchi:2013fk}. However,
this should not be surprising since TMAO is mostly hydrophobic and
thus might be better accommodated in the polymer hydration shell than
in bulk solution. But a key observation is that $K_p$ displays
significant conformation-dependence: it is higher near a collapsed
configuration than near an extended configuration. Therefore, TMAO
strongly interacts with the polymer ($K_p > 1$) but there is an
effective depletion next to extended conformations of polymer relative
to collapsed conformations.
  
  Although the local-bulk partition coefficient provides a better
  description of the preferential interaction because of its intensive
  nature, it is not directly connected to experimental observables. In
  contrast, the preferential binding coefficient $\Gamma$ can be
  measured experimentally, e.g. using the vapor-pressure osmometry
  technique \cite{Courtenay:2000uq}. The effect of preferential
  binding on a conformational equilibrium between the folded and the
  unfolded configurations $\mathrm{F}\rightleftharpoons \mathrm{U}$
  (with an equilibrium constant $K$) is usually understood in terms of
  the thermodynamic calculation first introduced by Wyman and Tanford
  \cite{WYMAN:1964vn,Tanford:1969kx}, which leads to
  \begin{equation}
  \frac{\partial \ln K}{\partial\ln a_s} = \Delta \Gamma_{F\rightarrow U} ,
  \label{eq:tanford}
  \end{equation}
  where $a_s$ is the activity of the cosolvent in the binary
  solution. According to Eq.~\ref{eq:tanford}, an increase in the
  concentration of the cosolvent would lead to the biomolecule
  unfolding if $\Delta \Gamma>0$, and in contrast would favor the
  folded state over the unfolded one if $\Delta \Gamma <0$. 

  In Figure~\ref{fig:pref_binding}C-D we show the average preferential
  binding coefficients $\Gamma$ for both collapsed and extended
  conformations in TMAO and urea solutions. As already discussed
  above, one should consider the $\Gamma$ values at the distance
  corresponding to the polymer first solvation shell. Not
  surprisingly, in all cases $\Gamma$ is positive, which is equivalent
  to $K_p>1$ (Figure~\ref{fig:pref_binding}A-B). Similarly, the trends
  in the difference between the extended and collapsed configurations
  $\Delta\Gamma$ follow that of $\Delta K_p$: $\Delta\Gamma$ is
  negative for TMAO, which stabilizes the folded state over the
  unfolded one, while the positive $\Delta\Gamma$ for urea clearly
  corresponds to its denaturing effect. Our results are therefore in
  agreement with the current consensus on the osmolyte
  effect\cite{Timasheff:2002kx,Parsegian:2000uq,bolen,Courtenay:2000uq,Canchi:2013fk}
  summarized by Eq.\ref{eq:tanford}.
  
  The main difference between our work and previous studies on
  proteins is that the sign of $\Delta\Gamma$ is different from that
  of $\Gamma$ for TMAO (they have the same sign for urea). This
  surprising observation can be understood if we consider the
  relationship between $K_p$ and $\Gamma$. Indeed combining
  Eqs.~\ref{eq:pref_bind} and~\ref{eq:pref_bind2} with the hypothesis
  that the bulk domain is large with respect to the local domain
  (i.e., $\langle n_X \rangle \ll N^{tot}_X$), leads to
  \begin{equation}
  \Gamma = \langle n_s \rangle \left ( 1-\frac{1}{K_p} \right ).
 \end{equation} 
 Therefore $\Delta\Gamma$ will depend on both $\Delta \langle n_s
 \rangle$ and $\Delta (1/K_p)$ (note that these two terms are not
 independent of each other). In experimental studies of proteins, it
 was found that $\Gamma$ is proportional to the solvent surface
 accessible area $S$\cite{Courtenay:2000uq}. Since $ \langle n_s
 \rangle$ is also proportional to $S$, $K_p$ is expected to be the
 same whether the protein is folded or not. This may not be surprising
 given that the nature of the exposed groups is the same in the folded
 and the unfolded state. However for our polymer, $K_p$ is very much
 conformation-dependent, while $n_s$ is only marginally higher in the
 extended state. This therefore provides an explanation for why
 $\Delta\Gamma$ is negative while TMAO always accumulates in the
 hydration shell ($\Gamma > 0$). Finally, it is interesting to note
 that simulations of decaalanine have found $\Gamma$ to be positive
 for TMAO\cite{Pettitt:2011ja} even though it was observed to behave
 as a protective osmolyte.

 \subsubsection{Conformation-dependence of the osmolyte chemical potentials in the polymer first solvation shell}

 To obtain a better understanding of how TMAO can preferentially bind
 to the polymer surface, yet still behave as a protective osmolyte
 favoring the polymer collapsed state, we investigate the free-energy
 changes (chemical potential) associated with the insertion of a
 single osmolyte or water molecule next to different conformations
 (both collapsed and extended) of the polymer.  These chemical
 potentials were determined from thermodynamic integration (see
 Methods, Appendix and Figures~\ref{fep_TMAO} and ~\ref{fep_Urea}). Insertion of a TMAO and a
 water molecule in the 1 M TMAO solution (or urea and water in the 7 M
 urea solution) was considered in three different cases: in the bulk,
 i.e. far from the polymer; in the first hydration shell of the
 polymer frozen in a \textit{collapsed} configuration; and in the
 first hydration shell of polymer frozen in an \textit{extended}
 configuration.

 Table~\ref{DeltaGtab} lists the results of thermodynamic integration;
 namely, the van der Waals and electrostatic contributions to the
 chemical potentials of urea and TMAO in the different cases.  In all
 cases insertion of a osmolyte molecule is more favorable next to the
 polymer than it is in the bulk: this is in agreement with the
 preferential binding values discussed above. However, the chemical
 potentials of TMAO and urea, relative to bulk values follow opposite
 trends as far as the conformation-dependence is concerned.  An
 inserted single TMAO molecule is more stable (has lower free energy)
 next to the \textit{collapsed} conformation of polymer than next to
 the \textit{extended}. Its free-energy is lower by 0.8~kcal/mol per
 TMAO molecule. This is mainly due to the more favorable free energy
 contribution from the van der Waals interaction, which overcomes the
 slight destabilization in the electrostatic contribution in the
 collapsed conformation due to less exposure to water or other TMAO
 molecules in the collapsed state than in the extended state. Given
 the importance of the van der Waals term for this (mostly)
 hydrophobic molecule, the total free-energy change is dominated by
 this contribution. In urea, however, the van der Waals term is small
 and does not totally compensate the electrostatic
 contribution. Therefore the insertion of a urea molecule is more
 favorable next to the extended state.

\begin{table*}[t]
\begin{center}
 \vspace{0.2 in}
 \begin{tabular}{lccccc} \hline System & $G_{vdw}$ & $G_{coulomb}$ &
$G_{total}$ & $\Delta \mu ^{bulk}$ & $<N_{k}>$  \\ \hline\hline TMAO &&&&\\ bulk & +1.99(0.02) & --13.97(0.02) &
--11.97(0.01) & 0 & --  \\ collapsed & --0.52(0.36) & --13.08(0.22) & --13.60(0.27) &
\textbf{--1.63} & 6.2(0.2) \\ extended & +1.04(0.45) & --13.80(0.08) & --12.76(0.38) & \textbf{--0.79} & 6.3(0.6) 
\\ \hline
 urea &&&&  \\  bulk & --0.17(0.04) & --13.27(0.05) &
--13.44(0.09) & 0 & --  \\ collapsed & --1.33(0.71) & --12.45(0.53) & ---13.78(0.20) &
\textbf{--0.33} & 28.6(0.3) \\ extended & --0.99(0.29) & --12.91(0.23) & --13.90(0.15) & \textbf{--0.46} &
41.9(0.8) \\ \hline
  \end{tabular}

  \end{center}
  \caption{Different free-energy contributions (in units of kcal/mol)
    for inserting single osmolyte molecules in the first solvation shell of the polymer (with
    $\epsilon_b=1.0$ kJ/mol) in 1 M TMAO and 7 M urea. $\Delta \mu ^{bulk}$ represents the difference with respect to the bulk solution, and the average number of molecules in first solvation shell of the polymer $N_{k}$ is also given. Standard deviations are given within parentheses.}
  \label{DeltaGtab} 
\end{table*}

The observed differences and respective contributions of van der Waals
and electrostatic free-energy, which drive the preferential
interaction with one state or another, suggest a mostly enthalpic
origin to this behavior. This is confirmed by considering the
distributions of both van der Waals and electrostatic energy
distributions for single osmolyte molecule in the hydration shell of
the polymer, either in a collapsed or in an extended conformation (see
Appendix and Table III). The observed trends are similar to that found
from the FEP study --- TMAO interacts preferentially with the polymer
collapsed state because of the van der Waals contribution,
leading to a difference of $\approx$~0.2~kcal/mol as compared to the
extended state and $\approx$~0.5~kcal/mol with respect to the bulk
phase. In contrast, a slight stabilization of urea in the extended
polymer hydration shell with respect to the collapsed state is found
($\approx$~0.2~kcal/mol) and appears to be driven by the electrostatic
contribution. In both cases, a significant stabilization is found as
compared to the bulk phase ($\approx$~1.5~kcal/mol).

\subsubsection{A free-energy based model  for action of protecting and
  denaturing osmolyte}

The discussion in the previous paragraphs have focussed on
only the contributions of the chemical potential of osmolyte (TMAO or
urea) to the free-energy difference between the polymer collapsed and
the extended states. We now try to build a free energy-based
thermodynamic model based on the above FEP data, with the goal of
validating it against the net PMF profiles of the polymer in the
respective solutions (as previously discussed in Figure
~\ref{pmf_all}). The net free-energy change for going from a collapsed
to an extended configuration can be expressed as
\begin{eqnarray}
\label{eq:delG}
\Delta G ^\mathrm{C-E}  &=   \Delta G_{gas}^\mathrm{C-E} &+  \sum \Delta G_{k}^\mathrm{C-E} \nonumber       \\
 &=  \Delta G_{gas}^\mathrm{C-E} &+  \Delta G_{w}^\mathrm{C-E} +  \Delta G_{cs}^\mathrm{C-E}\nonumber\\
                       & =  \Delta G_{gas}^\mathrm{C-E} &+  [N^{E}_w \mu ^E_w- N^{C}_w \mu ^C_w] \nonumber\\
                       &&+ [N^{E}_{s} \mu ^E_{s} - N^{C}_s \mu ^C_s]  
\end{eqnarray}
where $N^{C}_k$ (respectively $N^{E}_k$) is the average number of solvent
molecules of type $k$ (water $w$ or osmolyte $s$) in the first
hydration shell of the polymer in a collapsed (respect. extended)
configuration, and $\mu ^C_k$ (respect. $\mu ^E_k$) their associated
chemical potentials.  We assume here that the chemical potentials of
molecules beyond the first hydration shell are similar for both
polymer configurations.  

To evaluate Eq.~\ref{eq:delG}, we must separately determine three
individual contributions: 
 
(a) The first term of the equation corresponds to the free-energy difference between the
collapsed and extended states of the polymer itself in gas phase. It was obtained by repeating our
simulations in the gas phase and by performing umbrella sampling
calculations to estimate the free-energy difference between the
collapsed and the extended state in the absence of solvent, which was
found to be $\Delta G_{gas}^\mathrm{C-E} = 9.6$~kcal/mol in favor of the
\textit{collapsed} conformation. 
 
(b) To calculate the second term, which describes water's
  contribution to the total free energy difference, we repeated our
  FEP calculations for water as well using the same method as detailed
  above for the osmolyte molecules. Results are reported in
  Table~\ref{tab:DeltaGtab_w}. In all cases (bulk, TMAO or urea
  solutions), the difference in chemical potential between inserting a
  water molecule next to the collapsed or next to the extended state
  is very small, and slightly negative with respect to bulk. At the
  same time, the hydration number is significantly increasing because
  of a larger surface area in the extended state. Both the changes in
  chemical potential $\delta \mu _w = \mu^{E}_w - \mu^{C}_w$ and that
  of the number of molecules $\delta N_w=N^{E}_w - N^{C}_w$ contribute
  to the total free-energy difference between the collapsed and the
  extended state due to water molecules. A detailed analysis discussed
  in the Appendix and Table ~\ref{tab:decomp} shows that in all cases, the dominant contributions
  logically arise from $\delta N_w$. 
  
\begin{table}[t]
\begin{center}
 \vspace{0.2 in}

    \begin{tabular}{llcc} \hline System &&  $\Delta \mu ^ {bulk} $ & $<N_{k}>$ \\ \hline pure water &&&\\ &collapsed & 
--0.17(0.04) & 73.8  \\ &extended &  --0.15(0.02) & 111.3
\\  \hline TMAO &&&\\  &collapsed & 
 --0.08(0.05) & 66.9 \\ &extended &  --0.11(0.04) & 106.6
\\ \hline urea &&&\\ &collapsed & --0.13(0.14) & 33.0 \\ &extended & --0.17(0.31) & 43.5\\ \hline

  \end{tabular}

  \end{center}
  \caption{Total free-energy cost $\Delta \mu ^{bulk}$ (with respect
    to the bulk solution reference) for inserting single water
    molecules in the first solvation shell of the polymer (with
    $\epsilon_b=1.0$ kJ/mol) in pure water, 1 M TMAO and 7 M urea (in
    units of kcal/mol), and average number of water molecules in the
    polymer first hydration shell $N_{k}$. Standard deviations are given within
    parentheses.}
  \label{tab:DeltaGtab_w} 
\end{table}

(c) The third term describes contribution of osmolyte molecules
  (TMAO or urea) to the net free energy difference. The chemical
  potentials in each configuration were already discussed and are
  given in Table~\ref{DeltaGtab}, which also contains the total number
  of osmolyte molecules in the polymer first hydration
  shell. Similarly to water, we can define $\delta \mu _s = \mu^{E}_s
  - \mu^{C}_s$ and $\delta N_s=N^{E}_s - N^{C}_s$. For urea, both the
  $\delta \mu _s$ and $\delta N_s$ bring a negative contribution to
  the solvent-induced free-energy difference, which therefore favors
  the extended state. For TMAO, the two terms bring opposite
  contributions (Appendix and Table ~\ref{tab:decomp}). While $\delta N_s>0$ favors the extended
  conformation, it is the dominant contribution of the large $\delta
  \mu _s$ that leads to a positive $\Delta G_{s}^\mathrm{C-E}$,
  stabilizing the collapsed state relative to the extended one. 
 
\begin{figure}[t] \centering
\includegraphics[clip,width=1.0\columnwidth,keepaspectratio]{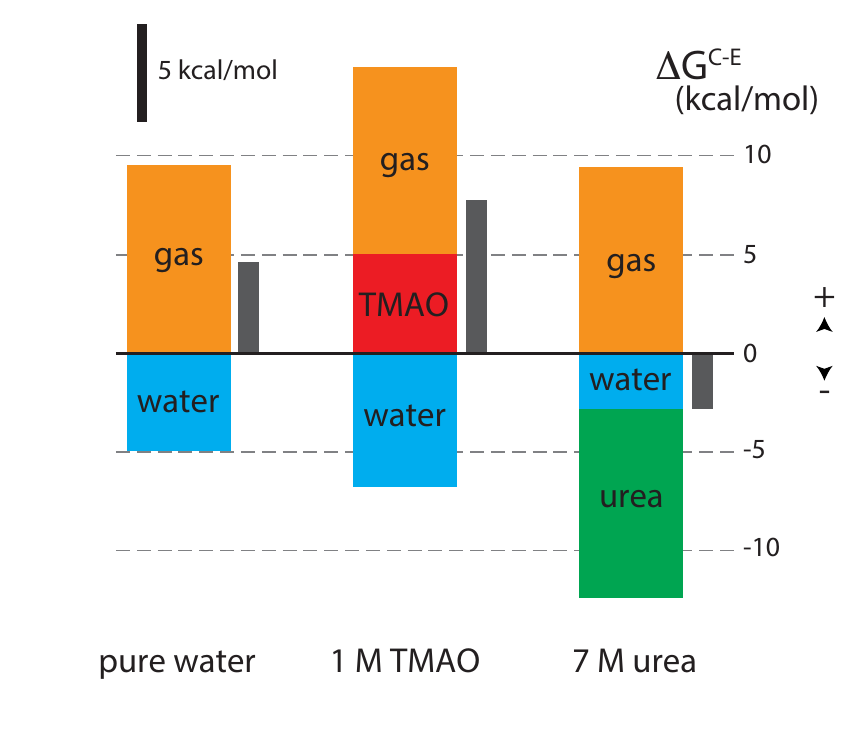}
\caption{Histograms showing the different contributions to the
  free-energy cost of inserting water or osmolyte molecules in the
  polymer hydration shell.}
\label{fig:deltaG}
\end{figure}

Figure~\ref{fig:deltaG} represents these three different individual
  contributions schematically. In pure water, the water contribution favors the
  \textit{extended} state, as shown above. However, it does not
  totally compensate  the polymer contribution (this will depend on the value of
    $\epsilon$), so that the folded state
  is still the most stable in this case, as calculated independently
  with our umbrella sampling simulations (Figure~\ref{pmf_all}). In
  the TMAO solution, the water contribution is only slightly perturbed with
  respect to the pure water case, but TMAO molecules will bring a
  contribution favoring the \textit{collapsed} state, in agreement
  with our earlier estimation based on PMF calculations
  (Figure~\ref{pmf_all}). Finally in the urea solution, the urea
  contribution stabilizes the \textit{extended} conformation greatly
  so that the overall free-energy difference becomes favorable towards
  extended conformations, as we have observed.

The semi-quantitative agreement that is observed in all three cases
with the free-energy differences calculated from umbrella sampling
simulations further validates the current approach. Note however that
for TMAO and urea solutions, we systematically overestimate the
respective stabilization. A possible explanation for this discrepancy
is that we consider only a few sites for insertion of a osmolyte or a
water molecule in the hydration shell. However, the fact that the
trend is correctly and self-consistently predicted and also that the
energy distributions exhibit exactly the same behavior makes us
confident in this approach. Finally, the decomposition presented in
the Appendix in terms of the respective contributions of $\delta \mu
_s$ and $\delta N _s$ also suggests that an osmolyte's behavior cannot
be predicted from the sign of $\delta \mu _s$ alone, as detailed in
the Appendix.\footnote{Indeed the main $\delta N_s$ contribution (which
  is equal to $\mu^{C}_s \delta N_s$) can overcome that of $\delta
  \mu _s$ if the relative variations of the hydration number when
  going from collapsed to extended conformations are larger than that
  of the associated chemical potentials, even if the osmolyte
  molecules are more stable around the collapsed conformations.}
  
\section{Conclusions}

In this paper, we considered the effect of osmolytes (TMAO and urea)
on a simple polymer, consisting of a short Lennard-Jones chain similar to an alkane
or lipid chain in aqueous solutions of the two osmolytes. This model
is reminiscent of the model used in our previous
paper\cite{Zangi:2009mi} directed at understanding urea denaturation
of hydrophobic collapse. 

Here we determined the free energy
landscapes of the hydrophobic polymer as a function of its radius of
gyration. We show that 1 M TMAO and 7 M urea act dramatically
differently on model polymer chains and their behaviors are sensitive
to the strength of the attractive dispersion interactions of the chain
with its environment: when these dispersion interactions are high
enough, TMAO suppresses the formation of extended conformations of the
hydrophobic polymer as compared to water, while urea promotes
formation of extended conformations. Quite surprisingly, we find that
\textit{both} protecting and denaturing osmolytes strongly interact
with the polymer (with both having a preferential binding constant
greater than zero), in contrast with existing explanations of the
osmolyte effect. An extensive free energy analysis suggests that
protective osmolytes are \textit{not} necessarily excluded from the
polymer surface. What really matters is the \textit{effective
  depletion} of the osmolyte as the polymer conformation
changes{, in agreement with the current consensus on the osmolyte effect\cite{Timasheff:2002kx,Parsegian:2000uq,bolen,Courtenay:2000uq,Canchi:2013fk}. Indeed for TMAO, it is the much more favorable free
energy of insertion of a single osmolyte near the collapsed
configurations of the polymer than near the extended configurations
that dictates its propensity to drive the system towards the collapsed
conformation, and therefore to lead to its protective effect. This
appears to be driven by van der Waals interactions. In contrast, urea is preferentially stabilized next to
the extended conformation because the smaller van der Waals contributions do not compensate the electrostatic contribution, suggesting this as an explanation of its
denaturing effect. 
   
A thermodynamic model taking into account the different contributions
(gas-phase, water and osmolyte) to the polymer conformational
equilibrium was developed. In the aqueous solution of urea, the free
energy contribution coming from urea and water easily cooperates to
shift the polymer towards extended conformations of the polymer. In
the aqueous solution of TMAO, TMAO's free energy contribution favors
the collapsed conformation and it is able to overcome the water's free
energy contribution which favors the extended conformation: overall,
the equilibrium is shifted towards the collapsed conformation.

We believe that this simple thermodynamic model provides an
  interesting perspective for explaining the role of protecting and
  denaturing osmolytes on simple macromolecules. Although the model is
  very simple, it provides fresh insights on the action of various
  osmolytes. Manipulation of simple polymers at the single molecule
  level has been recently achieved\cite{Li:2011fk} and we believe that
  the effect of osmolytes such as urea or TMAO on the polymer
  conformational equilibrium could be probed by such techniques. From
  a simulation perspective, it will be interesting in the future to
  extend our free-energy based approach to systems of increasing
  complexity like charged polymers, real peptides or proteins to shed
  light on the role of osmolytes on macromolecular conformations.

\section{Acknowledgments}
 This work was supported by grants from
the National Institutes of Health [NIH-GM4330 (to B.J.B.)] and by the
National Science Foundation through [via Grant No. NSF-CHE-0910943]. We gratefully acknowledge the computational support of the Computational Center for Nanotechnology Innovations (CCNI) at Rensselaer Polytechnic Institute (RPI).This work used the Extreme Science and Engineering Discovery Environment (XSEDE), which is supported by National Science Foundation grant number OCI-1053575.


\appendix

\section{Effect of the polymer bead parameter $\epsilon_b$ on the
  polymer conformational equilibrium}

To investigate how variations of the polymer bead parameter
$\epsilon_b$ affect its behavior in the various solutions we repeated
our simulations using 3 other values of $\epsilon_b$ (0.4, 0.6 and
0.8~kJ/mol). Note that the lower the value of this energy parameter
the more hydrophobic is the chain. Results are presented in
Figure~\ref{pmf_all_no_jacobi}.

In pure water the unfolded state gets stabilized relative to the
collapsed state as $\epsilon_b$ decreases (that is $\Delta
G_{F\rightarrow U}$ gets less positive) as $\epsilon_b$
decreases. When changing the $\epsilon_b$ parameter, there is a
competition between two opposite effects: first, as $\epsilon_b$
decreases this will decrease the interaction between beads and water
molecules, therebye increasing the hydrophobic character of the chain
and driving the polymer towards toward collapsed
configurations. However at the same time, the intramolecular
interactions will decrease, which would favor more extended
states. Our results suggest that this later contribution dominates.
In 1 M TMAO solution we observe that the protective effect
  is very sensitive to variations in $\epsilon_b$: indeed the
  protecting role evidenced above at $\epsilon_b=$~1~kJ/mol is
  moderate to almost nonexistent when the chain is really hydrophobic
(low values of $\epsilon_b=$ 0.4 and 0.6 kJ/mol), in agreement with a
previous study\cite{Athawale:2005kl}: the osmolyte effect becomes more
prominent only for $\epsilon_b=$ 0.8 and 1.0~kJ/mol. In strong
contrast with TMAO and in agreement with a previous
study\cite{Zangi:2009mi}, urea still readily denatures hydrophobic
polymers at small $\epsilon_b=0.4$~kJ/mol) yet its denaturing effect
becomes more prominent as $\epsilon_b$ increases.

 \begin{figure}\centering
\includegraphics[clip,width=0.8\columnwidth,keepaspectratio]{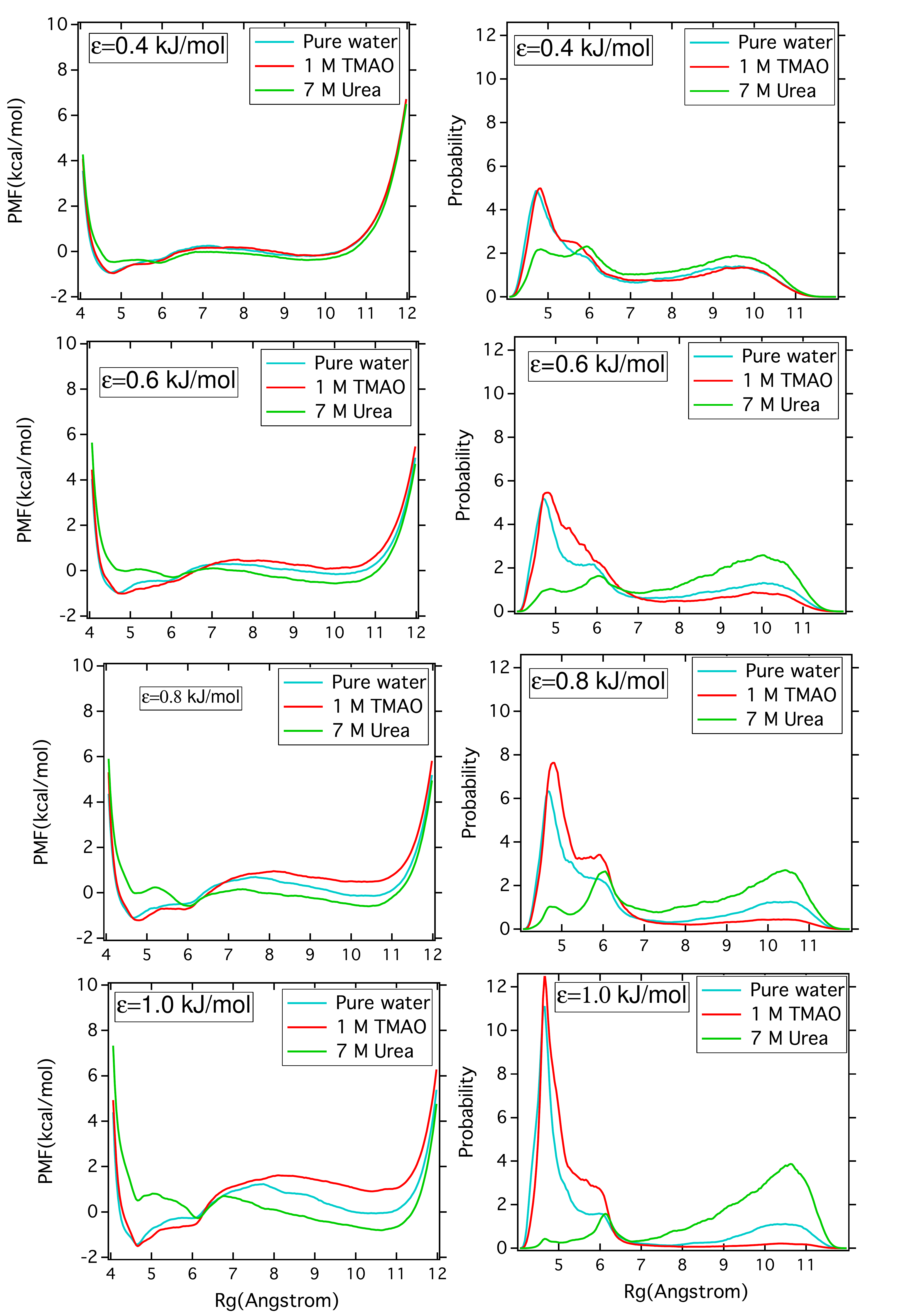}
\caption{PMFs along the radius of gyration (left) and the
  corresponding probability distributions (right) for four different
  polymer dispersion interactions ($\epsilon$= 0.4 kJ/mol, 0.6 kJ/mol,
  0.8 kJ/mol and 1.0 kJ/mol ) in three aqueous solutions (water, blue;
  TMAO, red and urea, green). The PMFs $W(R_g)$ are normalized so that
  $\int_0^\infty \exp(-W(r)/k_BT)dr=1$} \label{pmf_all_no_jacobi}

\end{figure}

\section{Effect of a different forcefield for TMAO on the free energy profile}

The force-field we have employed throughout our entire study for TMAO
(known as 'Kast model')\cite {kast} has also been recently modified by
Garcia and coworkers \cite{garcia} and a new forcefield, termed the
'osmotic model' has also been proposed by them. We repeated our
umbrella sampling simulations for free energy profile at 1 M using
this modified version of this force-field. As shown in the
Figure~\ref{PMF_Garcia}, its effect on the polymer chain is 
similar to that obtained from the original 'Kast model' of TMAO .

\begin{figure}[h!]  
\includegraphics[width=3.1in]{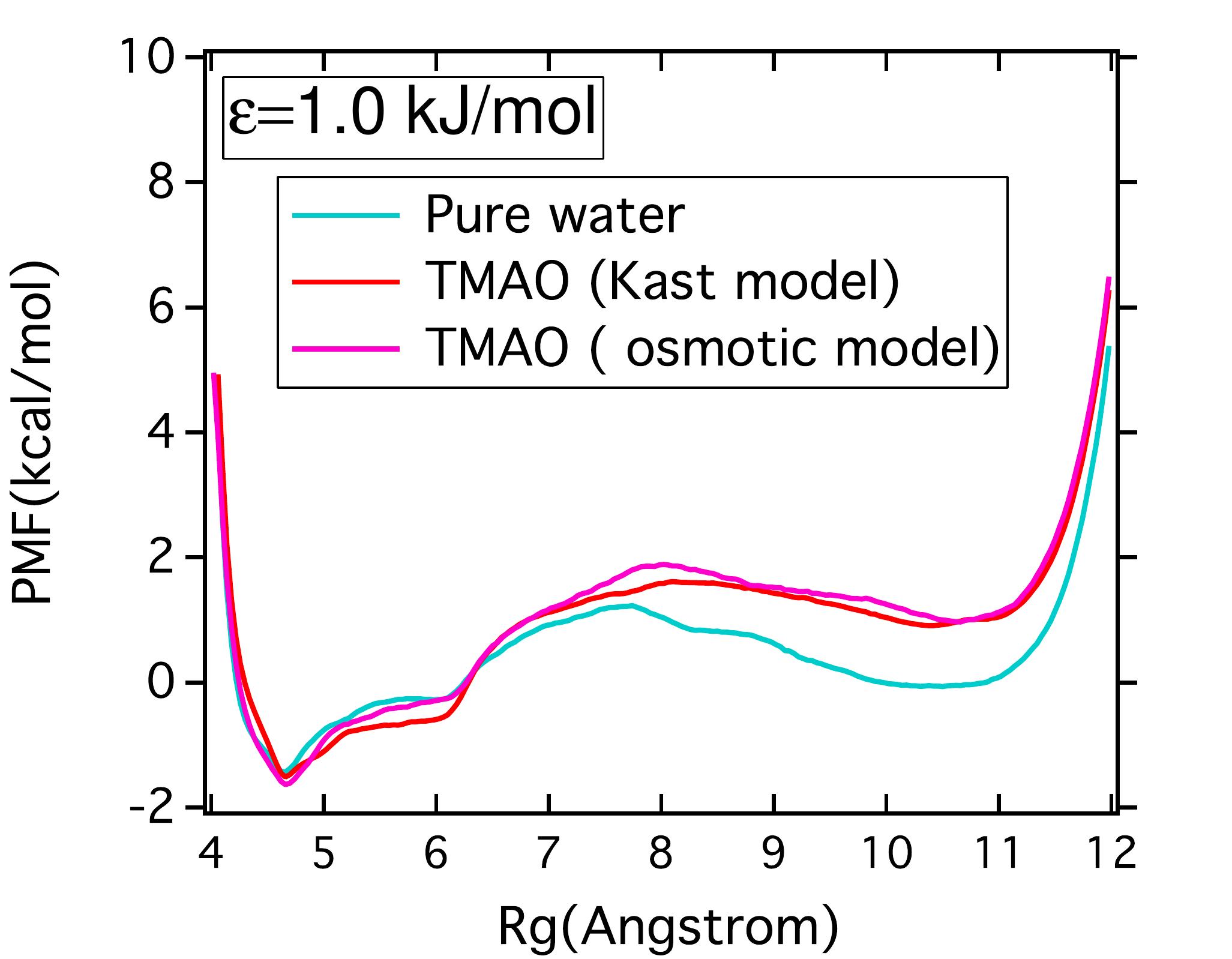}
\caption{Effect of different TMAO forcefields on free energy profile
  of polymeric chain in 1 M TMAO. `Osmotic model' is the modified
  version of TMAO forcefield as proposed by Garcia and co-workers and
  'Kast model' is the original TMAO forcefield (used in the rest of
  the current study). We also show the free energy profile of the
  polymeric chain in pure water.}\label{PMF_Garcia}
\end{figure}

\section{Effect of a different water model  on the free energy profile}

All the results reported in the main article were obtained using the
SPC/E water model\cite{Berendsen:JPhysChem:1987}. However, for the purpose of testing the
robustness of our results with respect to the different water models,
we have repeated our computation of free energy profiles using a
different water model, namely TIP4P water\cite{tip4p}, mainly because
it was used in our previous work on urea\cite{Zangi:2009mi}. Figure
~\ref{PMF_water_model} compares the effect of the two different water
models on the free energy profiles of the polymeric chain
($\epsilon_b=1.0$ kJ/mol) in different osmolyte solutions. The common
trend prevalent in all free energy profiles corresponding to TIP4P
water model is that the free energy barrier $\Delta G^{\ddag}_{F\rightarrow
  U}$ for the transformation from the compact to the extended state
in pure water and in 1 M TMAO are respectively higher for TIP4P water then
for SPC/E water.  For 7~M urea , the
extent of free energy of stabilization is negligibly more favorable in
SPC/E water  than in TIP4P water. But, interestingly,
the relative free energy difference $\Delta\Delta G_{F\rightarrow U}$ of the 
chain in going from  pure water to 1 M TMAO solution or 7 M urea
solution is almost the same for both SPC/E and TIP4P water models.

\begin{figure}[h!]  
\includegraphics[width=3.1in]{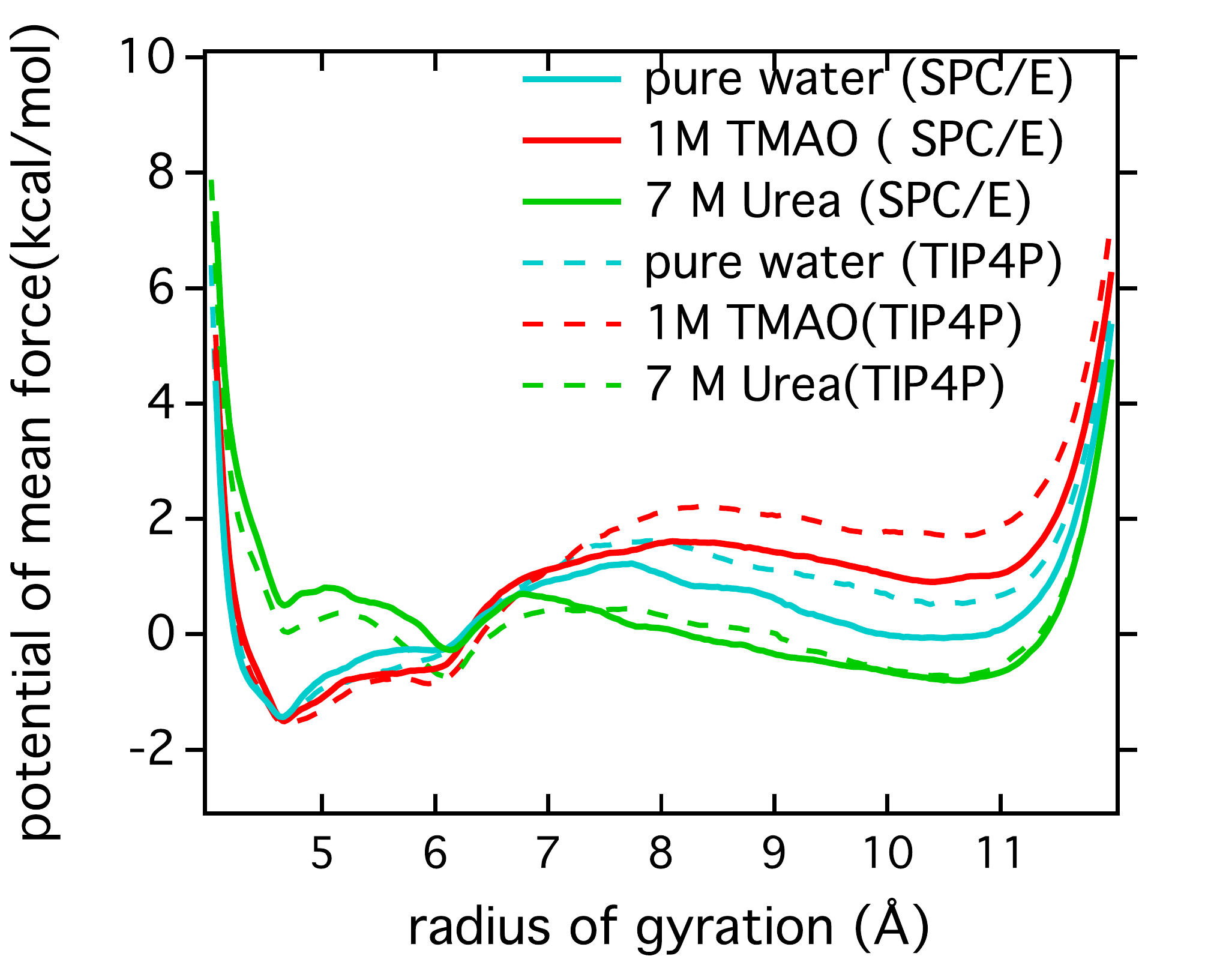}
\caption{Comparison of different free energy profiles of polymer chain ($\epsilon_b=1.0$ kJ/mol) obtained using SPC/E water model (solid lines) with that obtained from TIP4P water model (dashed lines).   }\label{PMF_water_model}
\end{figure}

\section{Thermodynamic integration}

As detailed in the Methods section, we performed free-energy
perturbation calculations to estimate the cost of inserting water or
cosolvent molecules at different positions in the polymer hydration
shell, or in the bulk. The interactions of the molecule being inserted
were slowly turned on in two stages: in the first stage only the van
der Waals interactions were turned on, and in the second stage the
electrostatic interactions were turned on. Thermodynamic integration
thus yields two contributions to the transfer free energy.  In
Figure~\ref{fep_TMAO}, we present results of thermodynamic
integration for the two stages and compare insertion of a single TMAO
at a given position in the bulk, in the hydration shell of the polymer
in a collapsed configuration, and in the hydration shell of the
polymer in an extended configuration. The thermodynamic data
confirms that van der Walls contribution to the free energy of
inserting a single TMAO is more favorable near a collapsed
conformation than near an extended conformation, and it more than compensates
the favorable electrostatic contribution near an extended conformation
than that near the collapsed conformation. In Figure~\ref{fep_Urea},
we present the same data for inserting a single urea molecule in the
urea solution, where the reverse trend is observed: the more
favorable electrostatic contribution to the free energy of inserting a
single urea near the extended conformation overcomes the favorable van
der Waals contribution to free energy of inserting a single urea near
the collapsed conformation.

\begin{figure}[h] \centering
\includegraphics[width=2.5in]{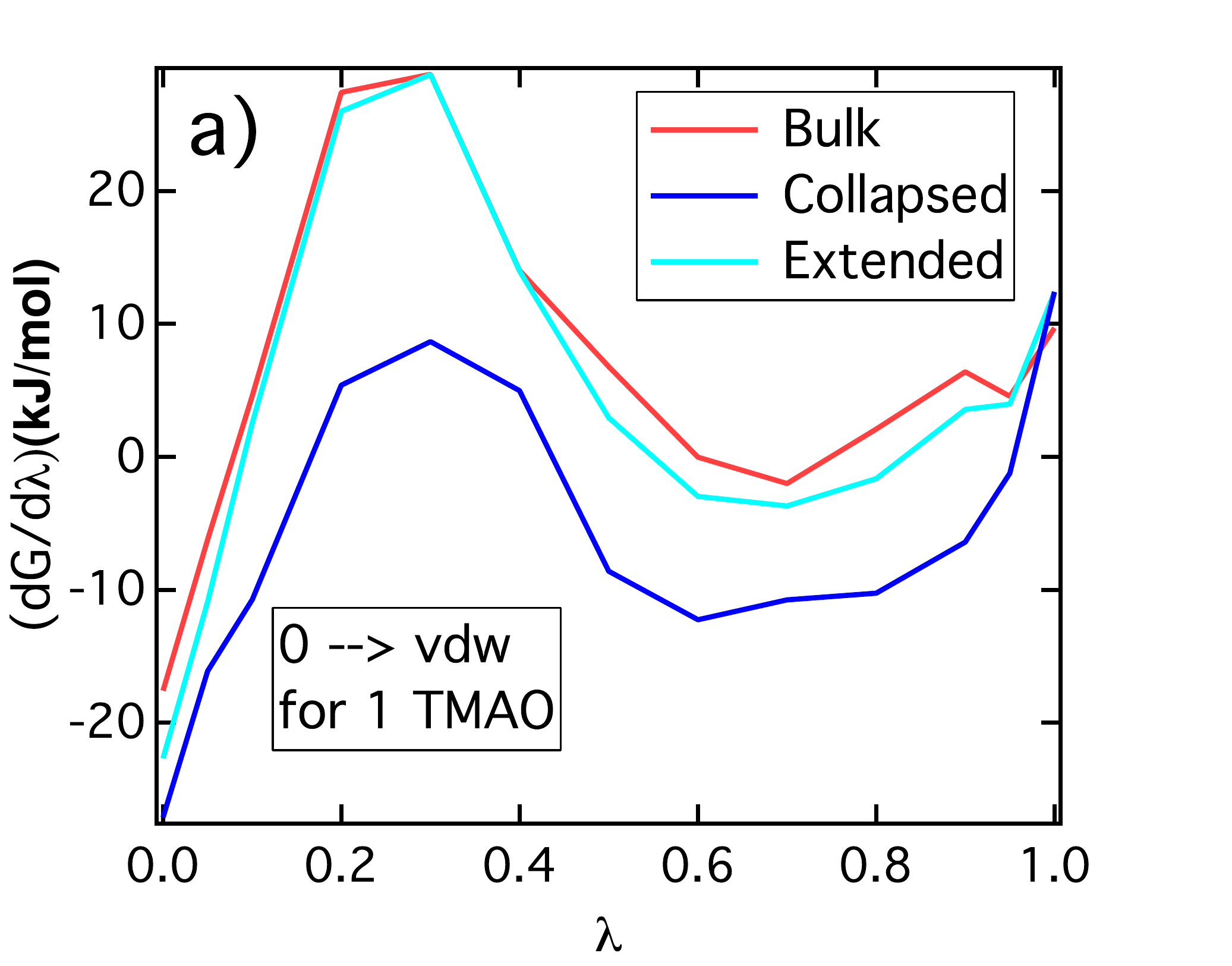}
\includegraphics[width=2.5in]{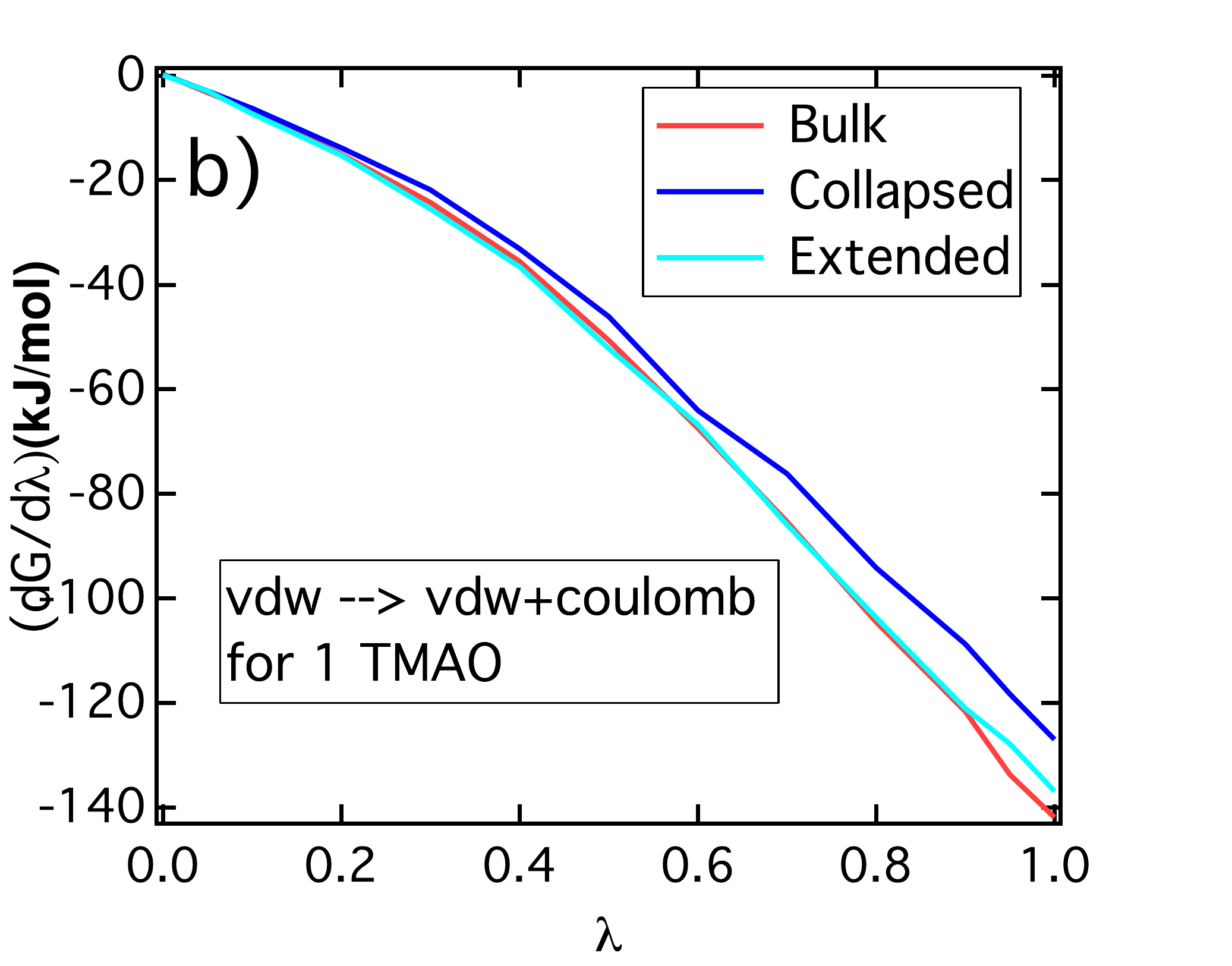}
\caption{Comparison of thermodynamic integration profile of
    turning on a) the van der Waals interactions and then b) the
    electrostatic interaction of a single TMAO molecule near the
    polymer for different polymer conformations in 1 M TMAO
    solution. }\label{fep_TMAO}
\end{figure}

\begin{figure}[h!]  \centering
\includegraphics[width=2.5in]{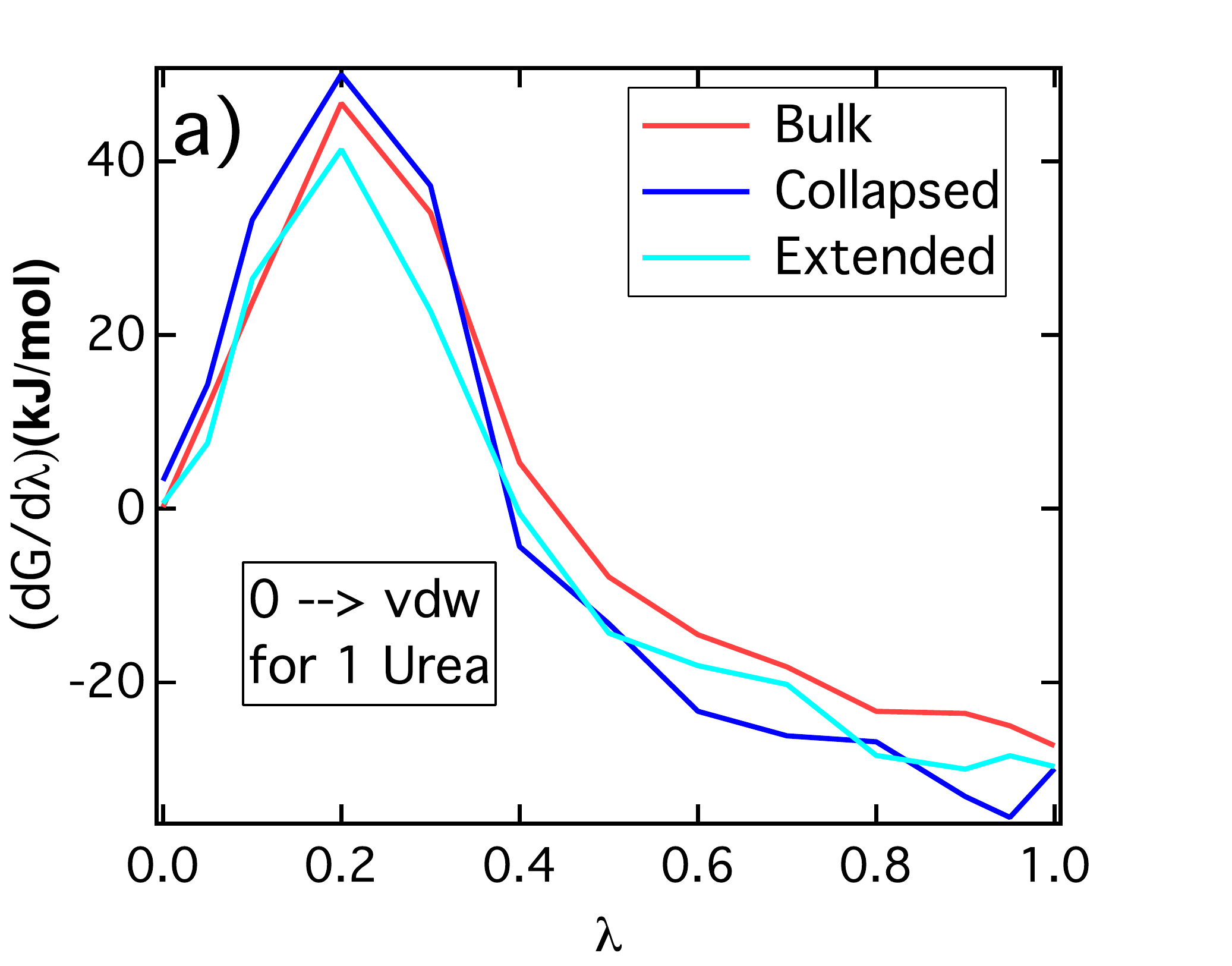}
\includegraphics[width=2.5in]{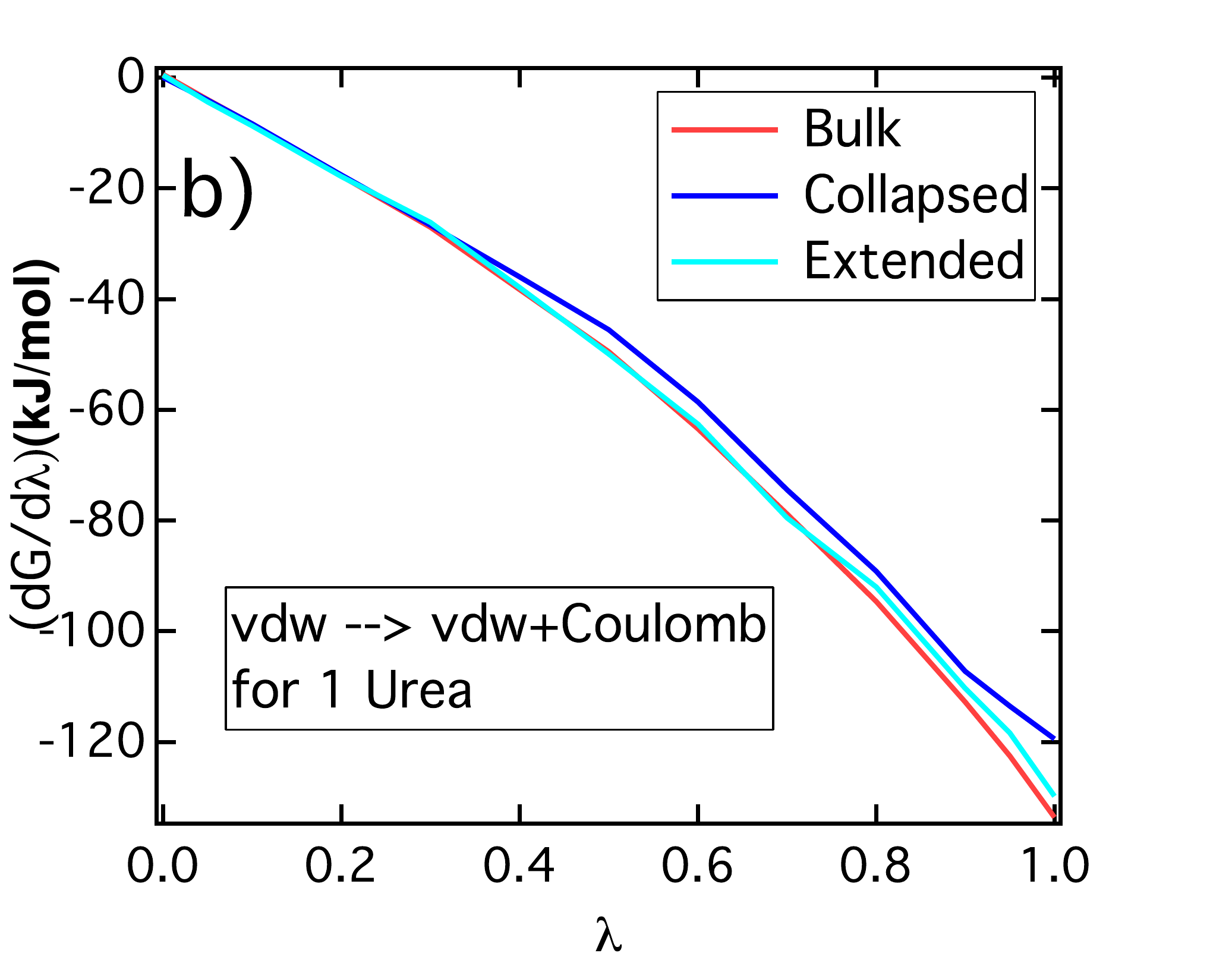}
\caption{Comparison of thermodynamic integration profile of turning on a)
the van der Waals interactions and b) the electrostatic
interactions of a single urea molecule near the
    polymer for different polymer conformations in 7 M urea solutions.}\label{fep_Urea}
\end{figure}

\section{Energy distribution in the polymer first hydration shell}

Table III  provides details on the respective net average
energies and the individual van der Waals and electrostatic
contributions to these energies for a single cosolute molecules (TMAO and urea). The
observed trends in the average energy analysis are consistent with
that found from the FEP study --- TMAO interacts preferentially with
the polymer collapsed state because of the favorable LJ contribution.
In contrast, a slight stabilization of urea in the extended polymer
hydration shell with respect to the collapsed state is found to be
driven by the electrostatic contribution.
\begin{table*}[t]
\begin{center}
 \vspace{0.3 in}
 \begin{tabular}{lccccc} \hline System & $E_{vdw}$ & $E_{coulomb}$ &
$E_{total}$ &  $\Delta E ^{bulk}$  \\ \hline\hline TMAO &&&\\ bulk & --3.824(2.22) & --29.510(4.71) & 
--33.334(3.82) & 0  \\ collapsed & --5.853(2.62) & --27.985(4.90) & --33.838(3.93) & \textbf{--0.504} \\ extended & --5.404(2.41) & --28.231(4.83) & --33.635(3.89) & \textbf{--0.301} 
\\ \hline
 urea &&&  \\  bulk & --5.079(2.55) & --28.234(5.16) &
--33.313(4.21) & 0  \\ collapsed & --8.451(2.54) & --26.329(4.90) & --34.780(3.98) &
\textbf{--1.467}  \\ extended & --8.002(2.61) & --26.993(4.89) & --34.995(3.99) & \textbf{--1.682}  \\ \hline
  \end{tabular}
   \caption{Different average energy contributions (in units of
    kcal/mol) of single cosolute molecules in the first solvation
    shell of the polymer in 1 M TMAO and 7 M urea. $\Delta E ^{bulk}$
    represents the difference with respect to the bulk
    solution. Standard deviations are given within parentheses.}

  \end{center} 
  \label{tab:DeltaEtab} 
\end{table*}

\section{Decomposition of the solvent contributions to polymer collapse}

In the pure solvent as well as in a mixture of water and a cosolute
(TMAO or urea), the contribution of the species $k$ to the free-energy
difference between the collapsed and the extended state can be
expressed as
\begin{equation}
\Delta G_{k}^\mathrm{C-E}= N^{E}_k \mu ^E_k - N^{C}_k \mu ^C_k
\label{eq:decomp}
\end{equation}
where $N^{C}_k$ (respectively $N^{E}_k$) is the number of solvent
molecules of type $k$ in the first hydration shell of the polymer in a
collapsed (respect. extended) configuration, and $\mu ^C_k$
(respect. $\mu ^E_k$) their associated chemical potentials. We can
define the relative changes for both quantities as $N^{E}_k = N^{C}_k
+ \delta N_k$ and $\mu^{E}_k = \mu^{C}_k + \delta \mu_k$. Substituting
$N^{E}_k$ and $\mu^{E}_k$ in Eq.~\ref{eq:decomp} leads after
simplification to
\begin{equation}
\Delta G_{k}^\mathrm{C-E}= N^{C}_k \delta \mu_k  + \mu^{C}_k \delta N _ k + \delta N_k \delta \mu_k
\label{eq:decomp_final}
\end{equation}
The first term on the right hand side of Eq.~\ref{eq:decomp_final}
represents the contribution of the change in chemical potential, for a
fixed number of molecules in the hydration shell; the second term
represents the contribution of the change in the hydration number for
a fixed chemical potential. The last one is a second-order term, whose
contribution is expected to be minor.

\begin{table}[t]
\begin{center}
 \vspace{0.2 in}
 \begin{tabular}{lcccc} \hline Case & $ N^{C}_k \delta \mu_k$ & $\mu^{C}_k \delta N _ k$ &
$\delta N_k \delta \mu_k$ & $\Delta G_{k}^\mathrm{C-E}$ \\ \hline\hline TMAO & \textbf{+5.2} & --0.2 & +0.10 &
+5.1 \\ urea & -3.5 & \textbf{--4.5} & --1.6 & -9.6 \\ 
water (bulk) & +0.8 & \textbf{--6.2} & +0.4& --5.0 \\
water (TMAO) & --2.4 & \textbf{--3.1} & -1.4 & --6.9 \\
water (urea) & --1.1 & \textbf{--1.4} & --0.4 & --2.9 \\ 
\hline
  \end{tabular}

  \end{center}
   \caption{Different free-energy contributions (in units of kcal/mol) of Eq.~\ref{eq:decomp_final}
  \label{tab:decomp} }
\end{table}

The different free-energy contributions for each solvent type is given
in Table \ref{tab:decomp}. In all cases, the chemical potential of a
water or cosolute (TMAO and urea) molecule in the hydration shell is
lower than in bulk, so that $\mu^{C}_k \delta N _ k$ is negative. This
represents the main contribution to the free-energy difference induced
by water (in all solutions) and urea, showing that the large change in
the hydration number is responsible for the increased solvent-induced
stability of the unfolded state over the collapsed one. The situation
is markedly different for TMAO. The large change in $\mu_s$ when going
from collapsed to extended is bringing a large $\delta N_s \delta
\mu_s$ contribution, which is positive and not totally compensated by
the $\mu^{C}_s \delta N _ s$ term (the $\delta N_s \delta \mu_s$ term
brings an additional positive contribution). Therefore the total
free-energy difference is positive, in strong contrast with what is
observed for urea and water.

The above decomposition also suggests that an osmolyte's behavior
cannot be predicted from the sign of $\delta \mu _s$ alone. Indeed the
main $\delta N_s$ contribution (which is equal to $\mu^{C}_s \delta N
_ s$) can overcome that of $\delta \mu _s$ if the following criteria
are met: (i) the stabilization in the polymer hydration shell with
respect to the bulk is large, but the variation between the collapsed
and the extended state is small ($\mu_s \ll 0$ and $\delta \mu _s
\approx 0$) ; (ii) the osmolyte can accumulate in the hydration shell,
i.e. $\delta N _ s$ is largely positive. In other words, this
situation can arise if the relative variations of the hydration number
when going from collapsed to extended conformations are larger than
that of the associated chemical potentials, even if the cosolute
molecules are more stable around the collapsed conformations.


%

\end{document}